# Objective Task-based Evaluation of Quantitative Medical Imaging Methods: Emerging Frameworks and Future Directions

**Yan Liu[1,†], Huitian Xia[1,†], Nancy A. Obuchowski[2], Richard Laforest[3], Arman Rahmim[4,5,6,7], Barry A. Siegel[3,8], Abhinav K. Jha[1,3,8,*]**

[1]Department of Biomedical Engineering, Washington University, St. Louis, MO, USA

[2]Quantitative Health Sciences, Cleveland Clinic, Cleveland, OH, USA

[3]Mallinckrodt Institute of Radiology, Washington University, St. Louis, MO, USA

[4]Department of Radiology, University of British Columbia, Vancouver, Canada

[5]School of Biomedical Engineering, University of British Columbia, Vancouver, Canada

[6]Department of Integrative Oncology, BC Cancer Research Institute, Vancouver, Canada

[7]Department of Physics & Astronomy, University of British Columbia, Vancouver, Canada

[8]Alvin J. Siteman Cancer Center, Washington University, St. Louis, MO, USA

[†]These authors contributed equally.

[*]Corresponding author; E-mail address: a.jha@wustl.edu

## Abstract

Quantitative imaging (QI) is demonstrating strong promise across multiple clinical applications. For clinical translation of QI methods, objective evaluation on clinically relevant tasks is essential. To address this need, multiple evaluation strategies are being developed. In this paper, based on previous literature, we outline four emerging frameworks to perform evaluation studies of QI methods. We first discuss the use of virtual imaging trials (VITs) to evaluate QI methods. Next, we outline a no-gold-standard evaluation framework to clinically evaluate QI methods without ground truth. Third, a framework to evaluate QI methods for joint detection and quantification tasks is outlined. Finally, we outline a framework to evaluate QI methods that output multi-dimensional parameters, such as radiomic features. We review these frameworks, discussing their utilities and limitations. Further, we examine future research areas in evaluation of QI methods. Given the recent advancements in PET, including long axial field-of-view scanners and the development of artificial-intelligence algorithms, we present these frameworks in the context of PET.

**Key points:**

- There is an important need for frameworks for rigorous and objective task-based evaluation of quantitative imaging (QI) methods.
- We outline emerging frameworks for evaluating QI methods, including virtual imaging trials, no-gold-standard evaluation, evaluation for joint detection-quantification tasks, and evaluation for quantifying multi-dimensional parameters.

- We provide a critical review of these evaluation frameworks, discussing their utilities, limitations and areas of future research.

**Synopsis:**
QI holds significant potential across diverse clinical applications. For clinical translation of QI, rigorous and objective evaluation on clinically relevant tasks is essential. This paper outlines four emerging evaluation frameworks, including virtual imaging trials, evaluation with clinical data in the absence of ground truth, evaluation for joint detection and quantification tasks, and evaluation of QI methods that output multidimensional parameters. These frameworks are presented in the context of recent advancements in PET, such as long axial field of view PET and the development of AI algorithms for PET. We conclude by discussing future research directions for evaluating QI methods.

# I. Introduction

Quantitative imaging (QI), which involves extracting numerical or quantifiable features from medical images to assist with clinical decision-making, has shown promise across multiple clinical applications[1,2]. Examples include the use of metabolic tumor volume (MTV) and total lesion glycolysis (TLG) derived from positron emission tomography (PET) images to predict cancer treatment outcomes[3]; apparent diffusion coefficient measured using diffusion-weighted magnetic resonance imaging to monitor cancer therapy response[4]; myocardial blood flow measurement using quantitative PET to diagnose coronary artery disease[5]; whole-body PET parametric imaging to enhance lesion detectability and assist in differentiation between benign and malignant signal[6], and radiotracer uptake in lesions and radiosensitive organs quantified from single-photon emission computed tomography (SPECT) and PET images for treatment planning of targeted radionuclide therapy[7]. Further, radiomic features extracted from medical images have shown promise for personalized treatment[8]. QI is thus emerging as a highly promising paradigm for imaging-based clinical decision-making.

To realize the promise of QI, measurements yielded by the QI methods need to be evaluated for accuracy and precision[9]. A QI method that yields inaccurate measurement may not correctly reflect the underlying pathophysiology, and imprecise measurements may provide limited confidence in making clinical decisions. However, while important, evaluation of QI methods on quantification tasks presents multiple challenges. Such evaluation requires knowledge of ground-truth data, which may often be unavailable in clinical settings. Also, conducting such evaluation studies clinically may be expensive, time consuming and logistically challenging. Moreover, typically performing quantitative tasks in medical imaging first requires performing a detection step, and in those cases, an evaluation that is solely focused on evaluating performance on the quantification task may not be sufficient. Further, evaluating QI methods that output multidimensional parameters, such as those for radiomics, may require different strategies from those used for evaluating methods that output unidimensional parameters. In light of these challenges, there is an important need for frameworks to evaluate QI methods.

To address this important need, multiple techniques for task-based evaluation of QI methods are being developed. In this article, we review a subset of these techniques with the goal of outlining four emerging frameworks for performing such evaluation (Fig. 1). We first present the paradigm of virtual imaging trials to evaluate QI methods using cost-effective and efficient approaches. We then discuss the evaluation of QI methods using patient data without ground truth. Next, a framework to evaluate QI methods for joint detection and quantification tasks is outlined. Finally, we describe a framework to evaluate QI methods that output multi-dimensional parameters such as radiomic features. This article presents these evaluation frameworks in the context of PET. Recent advances in PET,, including long axial field of view (LAFOV) PET[10], radiomics[8], and the use of artificial intelligence (AI)-based algorithms for tasks such as image reconstruction, image enhancement and image segmentation[11], highlight the need for robust evaluation frameworks to support clinical translation of these advances, motivating the PET-centered articulation of these frameworks.

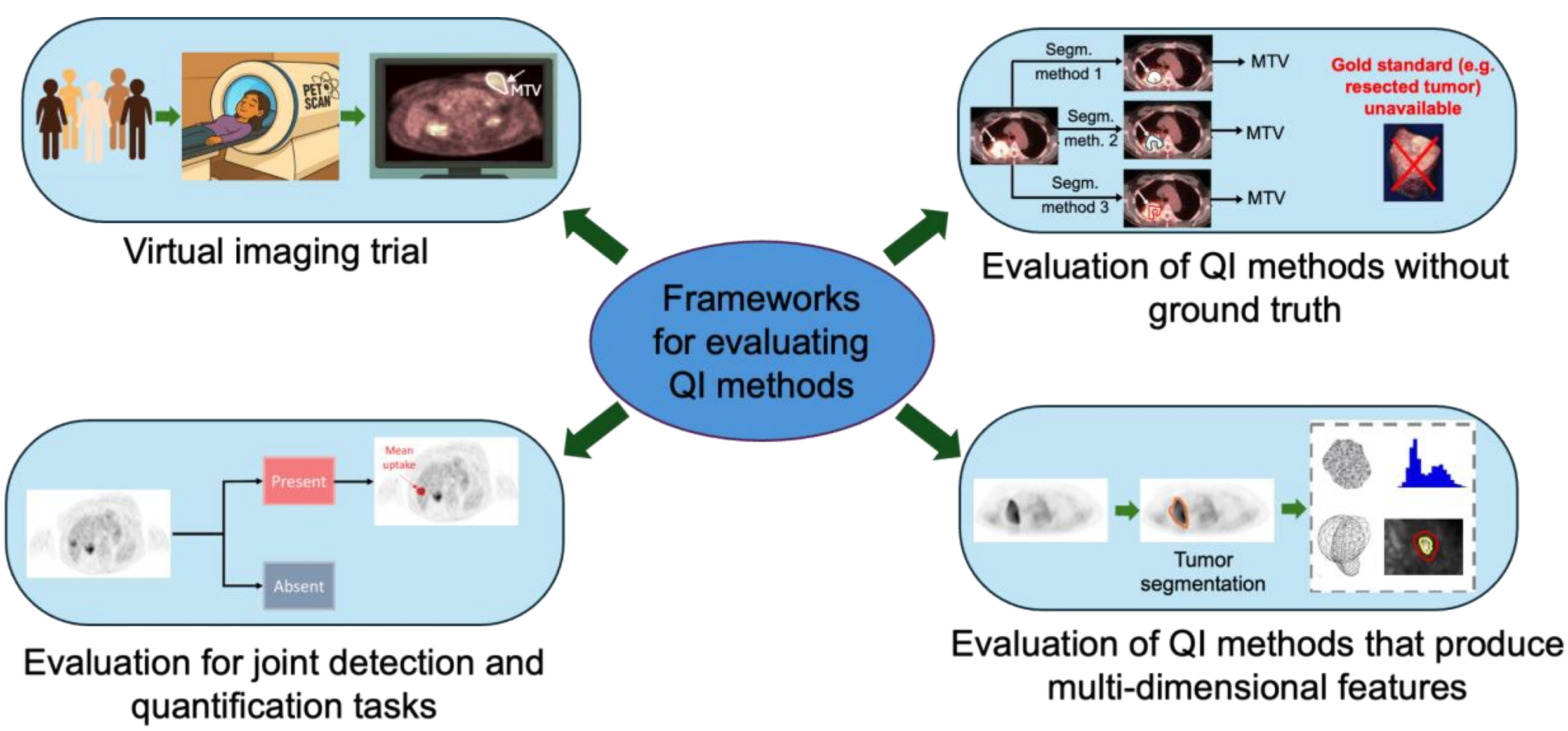

Fig. 1: An illustration of the frameworks for evaluating QI methods outlined in this article.

## II. Virtual Imaging Trials

Ideally, QI methods should be evaluated clinically with patient data. However, this presents multiple challenges including high costs, difficulties in recruiting subjects, increased time requirements, ethical concerns, radiation-exposure-related risks, and lack of ground truth for comparison[12]. Given these challenges, mechanisms are needed to identify promising methods for further clinical evaluation. The emerging paradigm of virtual imaging trials (VITs)[12,13] is showing promise in fulfilling this role. To evaluate QI methods, VITs with patient population and imaging systems replaced by a digital phantom-based clinically realistic population and a simulated scanner that accurately models the imaging physics, respectively, can be considered, as shown in Fig.2.

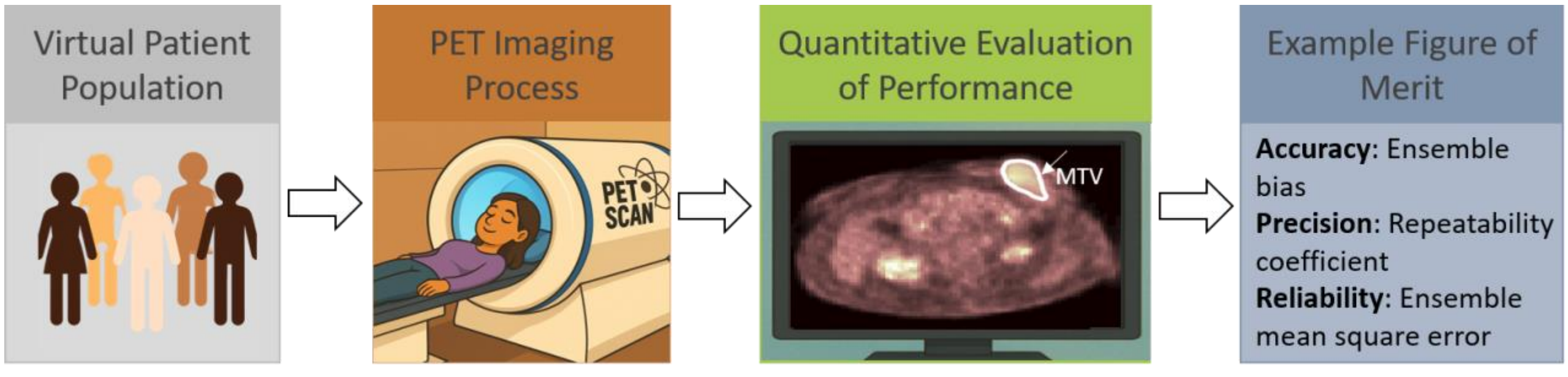

Fig. 2: A block diagram that demonstrates the process of conducting virtual imaging trials in the context of evaluating QI methods for PET, demonstrating the modeling of virtual patient population, imaging process, extraction of quantitative measurements and quantifying performance of the QI method using suitable figures of merit.

**Modeling Virtual Patient Population**: The first step in VITs is to simulate a clinically relevant virtual patient population. This requires accurate modeling of *in vivo* anatomical and physiological properties and variabilities in these properties in patient populations. High-resolution digital anthropomorphic phantoms, such as the extended cardiac and torso (XCAT) phantom[14], have been developed for this purpose. The XCAT phantom models anatomic variabilities, different sources of patient motion, and can be used to simulate physiological variabilities[14–16].

Realistic modeling of signals of interest, such as lesions, is also important. To generate realistic digital lesion phantoms, stochastic approaches have been developed for modeling parameters such as size, shape, location, and signal-to-background ratio[17]. Further, images generated by these approaches have been used to evaluate AI-based PET segmentation methods[17,18].

**Modeling Imaging System**: The next step in VITs is generating projection data corresponding to the digital patients. In PET, this can be accomplished through software that models the PET system instrumentation, including detector response, spatial, timing and energy resolution, and the positron and photon propagation, including photon-non-collinearity, positron range, attenuation, scatter, random coincidences and noise. To simulate PET systems, multiple software tools have been developed. A tabular representation of these tools, along with their trade-offs, is provided in Jha and colleagues[9]. These software tools can be categorized into Monte-Carlo (MC) approaches, such as GATE[19] and SIMSET[20], and analytical approaches such as SMART PET[21], ASIM[22], and FASTPET[23]. MC approaches, while providing accurate system modeling, are computationally expensive and time consuming, which may limit their feasibility for large-scale evaluation studies. Analytical approaches are computationally less expensive but may have lower accuracy. Thus, the choice of the simulation approach can be guided by the desired accuracy versus computational constraints.

**Performing the Quantification Task**: In medical imaging, including in PET, quantitative parameters are typically estimated from reconstructed images. To perform PET reconstruction in research studies, software such as STIR[24],CASTOR[25], and PyTomography[26] are widely used. Following reconstruction, in several cases (e.g., MTV or mean uptake quantification), a delineation of the region of interest is performed on the image, and the parameters are estimated from the delineated region. Another approach for quantification is to estimate the unknown parameters directly from projection data (projection-domain quantification)[27]. This approach has been primarily studied for uptake quantification[27,28], but also, more recently in quantifying parameters such as signal size and location[29].

From an estimation-theory perspective, quantification methods can generally be categorized into frequentist and Bayesian approaches[9]. In frequentist methods, the unknown parameter is assumed fixed. A common method in this category is maximum-likelihood estimator. In Bayesian methods, the unknown parameter is considered as a random variable, where a prior distribution of the parameter is assumed. The choice of cost functions in these approaches can impact the derived estimation method. Widely used estimators in this category include maximum-a-posteriori estimator and the posterior-mean estimator. As AI algorithms continue to advance QI, deep-learning-based quantification techniques have also emerged such as for quantifying standard uptake value ratio (SUVr) in amyloid PET[30]. Another recent study proposed a deep-learning-based posterior-mean estimator to quantify the fractional volume that a tumor occupies in each voxel in an oncological PET image[17].

When performing quantification tasks, it is important to be mindful of the concept of estimability[9].

A parameter is considered estimable if it could be accurately estimated for all possible true values from the measured image data. Caution should be exercised when estimating parameters that may not have been measured by the imaging system, such as estimating high-frequency features from a low-resolution PET image.

**Figures of Merit to Quantitatively Evaluate Performance**: The knowledge of ground truth in VIT facilitates objective evaluation of QI methods. Commonly used criteria for evaluation along with corresponding FoMs are listed below:

- **Accuracy:** The degree of closeness between the average measured values and the true value, typically quantified by measurement bias. If the bias depends on the true value, the use of bias profile, which quantifies the bias over different QI value ranges, is more desirable. Another option is ensemble bias, defined as the bias averaged over the distribution of true values. This has the advantage of being a summary FoM, although the true value distribution should be carefully specified.
- **Precision:** This quantifies the variability of different measurements under the same experimental conditions and includes criteria such as repeatability and reproducibility. Repeatability refers to precision under identical or nearly identical experimental conditions. Common FoMs include within-subject variance and repeatability coefficient. Reproducibility refers to precision under different conditions, including different clinical sites, scanners, operators or other factors, and can be quantified using the reproducibility coefficient. The impact of imaging-system-introduced noise can be quantified by the variance of multiple realizations given the same phantom and imaging system.
- **Overall reliability:** A summary FoM that incorporates both accuracy and precision while accounting for the variability in the estimated parameters, often quantified by mean squared error (MSE) and ensemble mean squared error (EMSE). EMSE averages the mean squared error over the distribution of the true values and noise in the imaging process.

The choice of FoM can be application dependent. For example, consider the assessment of regional SUVr quantified from amyloid PET images as a biomarker for amyloid positivity. Similar bias in the estimated values for the amyloid-positive and amyloid-negative patients may not be a concern. Instead, the precision of estimated values, which subsequently impacts the diagnostic task of separating these patient populations, is more relevant. Thus, in this case, precision-related FoMs may be more appropriate.

**Areas of Future Research**: VITs present a cost-effective, safe, and feasible evaluation approach to identify promising QI methods for further clinical evaluation. Given these advantages, further research is warranted to improve the accuracy and clinical realism of these VITs, particularly in the context of PET. One important area is modeling the temporal components of the radiotracer distribution in the digital phantoms. The distribution of the PET radiotracer within a patient's body is a dynamic process, wherein uptake varies over time based on physiological parameters, such as tissue perfusion and metabolism. Another important area of research is the acceleration of MC approaches to simulate PET systems. Recent studies have demonstrated the use of graphic processing units for MC approaches in PET[31,32], achieving orders of magnitude speed-up compared to conventional MC approaches. This provides an encouraging direction to accelerate PET system modeling without compromising accuracy.

Developing strategies for validation of VITs is another important research frontier, since VITs may have inherent limitations in fully capturing the complexities of real-world conditions. One approach is to assess realism of images generated in VITs, and to this effect, methods have been developed[33,34], but further research is needed. Another strategy is to develop methods to validate VITs on the eventual clinical tasks[35]. The verification, validation and uncertainty quantification (VVUQ) approach[36,37], may provide a methodological basis to assess VITs. VVUQ, when applied to VITs, helps ensure the underlying models solve the equations correctly (verification), accurately represent the real-world system being emulated (validation), and characterize the uncertainty in VIT predictions (uncertainty quantification) in the context of specific tasks[38]. This helps emphasize the relevance and reliability of VIT outputs.

## III. Evaluating quantitative imaging methods without ground truth

To generate evidence for clinical translation of QI methods, evaluation studies that demonstrate the efficacy of the methods with clinical data are needed. However, such evaluation typically requires knowledge of the ground truth, which is generally unavailable clinically. Toward addressing this issue, strategies have been developed to evaluate QI methods with and without reference standards[39], where a reference standard is defined as a well-accepted or commonly used method for measuring the quantitative value but with recognized bias and/or measurement error[39]. Given the error present with a reference standard, evaluation with a gold standard is more desirable, where a gold standard is defined as the best possible method to measure the quantitative value, and is ideally presumed to be correct for the parameter being estimated. However, obtaining such gold-standard data can be resource-intensive, time-consuming, and even impossible in many situations. Evaluation of QI methods with patient data has thus been hindered by the lack of suitable gold standards, as illustrated in Fig. 3.

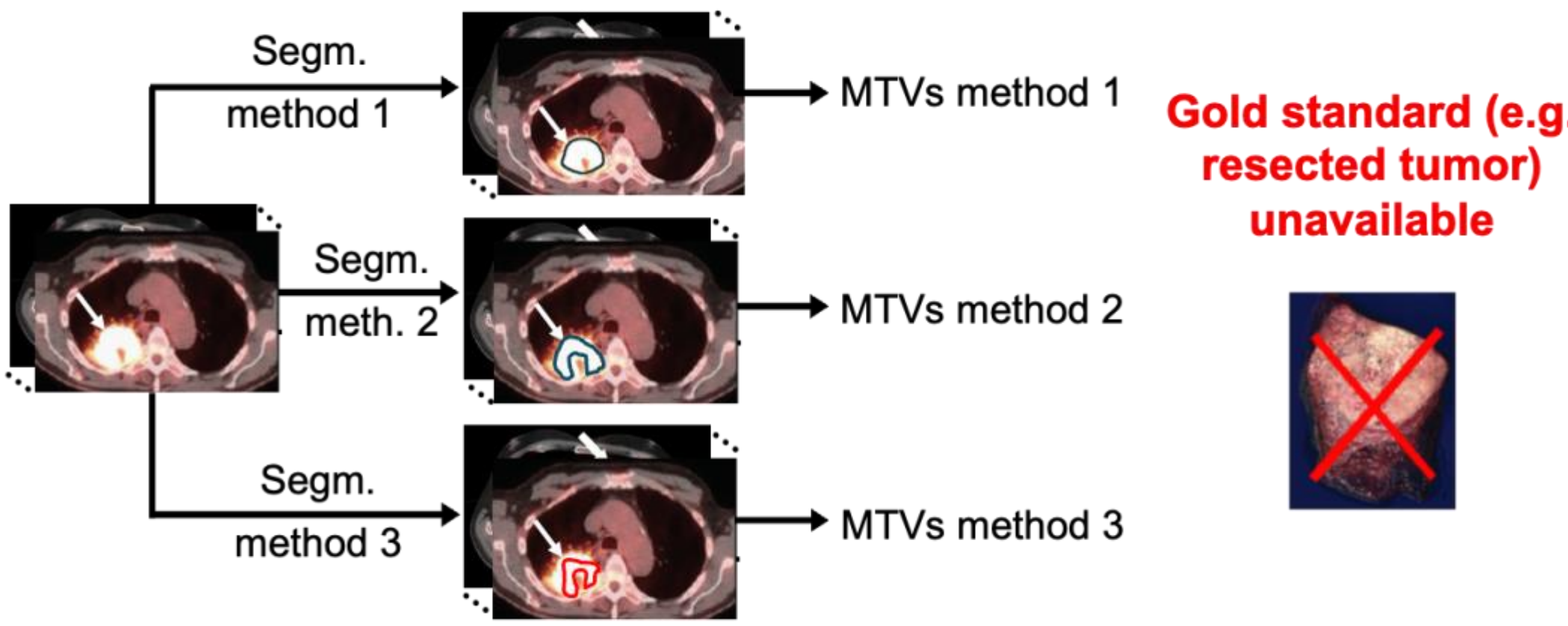

Fig. 3: Illustration of the challenge posed by the lack of a gold standard in evaluating three PET image segmentation methods for MTV quantification. The white arrow indicates the lesion.

To evaluate QI methods in the absence of ground truth, in this article, we discuss a class of no-gold-standard evaluation (NGSE) techniques that builds upon the regression-without-truth (RWT) technique[40,41]. These techniques are based on the premise that, since the measured values are the result of a specific image formation and quantification process applied to the true values, a relationship between the true and measured values is expected. More specifically, the RWT technique assumes a linear relationship between the measured and true values, characterized by a slope, a bias and the standard deviation of a zero-mean Gaussian noise term. Assuming that the true values are sampled from a parametric distribution with known bounds and that noise of different QI methods is uncorrelated, the RWT technique uses a maximum-likelihood approach to estimate the linear-relationship parameters without access to the true values. The ratio of the noise standard deviation to the slope, termed as noise-to-slope ratio (NSR), is then used as a FoM to evaluate different methods based on precision, with smaller NSR values indicating a more precise QI method.

The efficacy of the RWT technique was demonstrated in comparing software packages in cardiac SPECT[42] and segmentation methods in cardiac cine MR imaging[43] for measuring ventricular ejection fraction. The RWT technique was advanced to account for cases where the bounds of the true value distribution are unknown[44]. The resultant NGSE technique demonstrated efficacy in evaluating segmentation methods in diffusion MR imaging for estimating apparent diffusion coefficient[45] and reconstruction methods in SPECT for estimating mean regional activity uptake[44]. The NGSE technique was further advanced to model correlated noise of different QI methods[46]. The resultant technique demonstrated efficacy in ranking segmentation methods in PET for quantifying MTV[47] and ranking SPECT reconstruction methods for estimating regional activity uptake in alpha particle radiopharmaceutical therapy[48].

A framework for applying the NGSE technique was proposed and applied to evaluate different lesion segmentation methods in PET[49]. Here, we present this framework for evaluating several candidate QI methods using clinical data, as illustrated in Fig. 4, with key components summerized below.

**Check Linearity Between True and Measured Values**: The NGSE technique assumes a linear relationship between measured and true values for each QI method. This assumption can be verified through inter-method comparisons, realistic simulations, and phantom studies[49]. If linearity holds, the NGSE technique can be applied.

**Apply the No-Gold-Standard Evaluation Technique:** The input to the NGSE technique is the measured values from all QI methods. The technique estimates the NSR for each QI method, which serves as a FoM to rank the QI methods on the basis of precision of the measured quantitative values.

**Perform Consistency Checks:** Since the NGSE technique is a statistical procedure, there may be errors in the estimates yielded by this technique. Consistency checks can help flag such potential failures[42,49]. For instance, one such check compares whether NGSE-predicted linearity between QI methods match with actual data.

**Provide Rankings of the QI Methods:** The NSR values provide a FoM to rank the QI methods on the task of precisely measuring the quantitative value. However, these values are estimated from only one set of patient data, which rep- resents only a subset of the patient population. To account for uncertainty arising from population sampling, a bootstrap-based approach has been proposed to determine the ranking by computing the confidence intervals of the differences in estimated NSR values between each pair of methods[49]. For example, when comparing methods A and B, if the upper limit of the one-sided confidence interval for NSR_A− NSR_B is less than zero, we infer that method A is more precise than method B.

**Areas of Future Research**: The NGSE technique involves estimating multiple parameters that define the relationship between the true and measured values. Estimating these parameters may require multiple patient images[44], limiting the use of this technique to scenarios with large patient datasets. One potential approach to address this issue is incorporating prior information about the linear-relationship parameters, as can be obtained from realistic simulations and physical-phantom studies. This can then help reduce the required number of patient images[50]. Extending the NGSE framework to acquire and integrate such prior information is an important future direction. Additionally, while the NGSE technique has been validated in various clinical applications, these validations rely on simulations due to the need for ground truth. Evaluation with clinical gold standards would strengthen confidence for clinical translation of the NGSE technique.

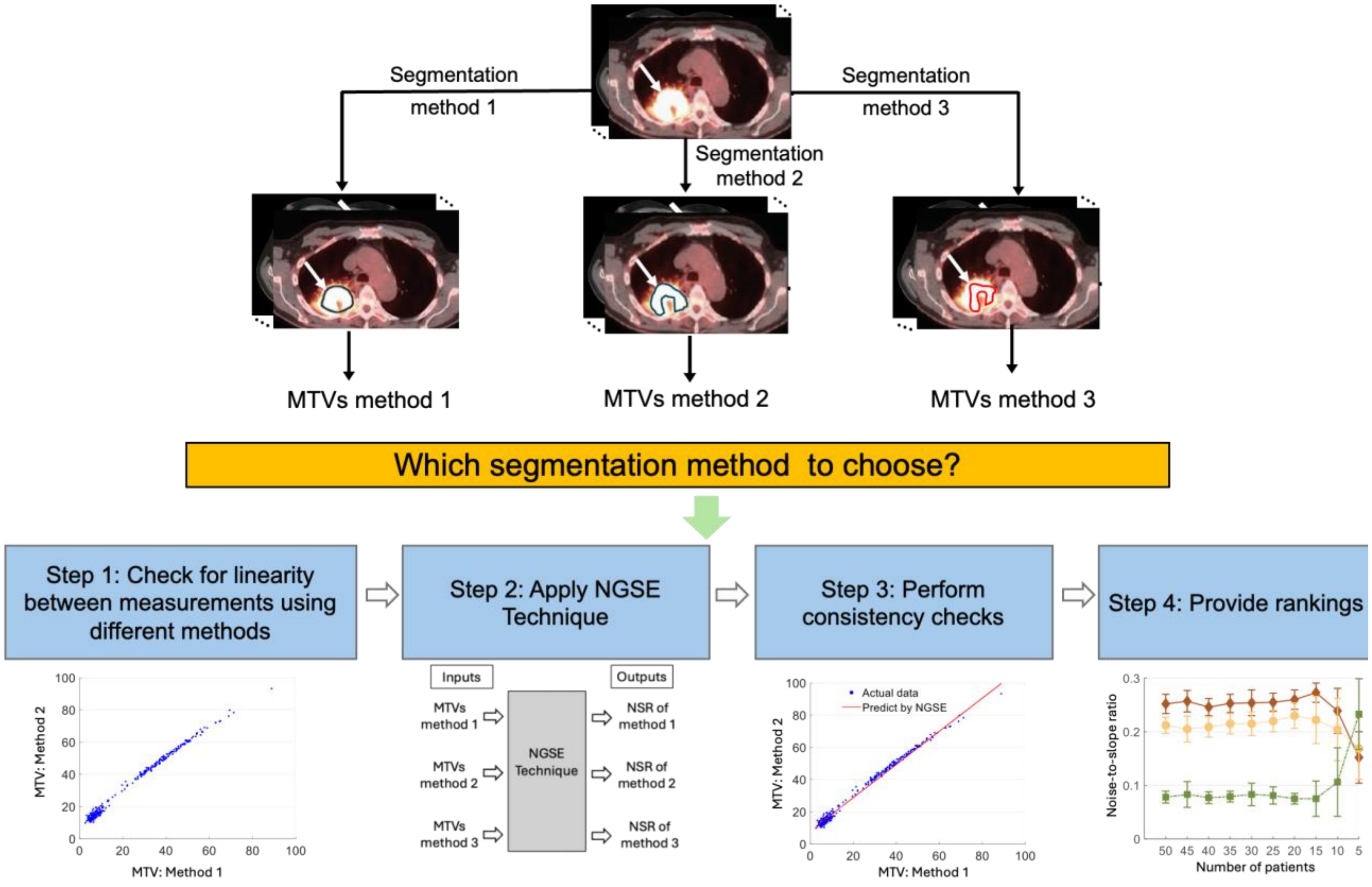

Fig. 4: A practical no-gold-standard evaluation framework to evaluate QI methods, illustrated by evaluating three PET image-segmentation methods on the task of quantifying MTV. The white arrow indicates the lesion.

## IV. Evaluation for Joint Detection and Quantification Tasks

Clinically, performing any quantification task typically requires a detection step. For example, to quantify MTV and TLG from a PET image, the first step is for a physician to visually interpret the image to detect the lesion(s) (Fig. 5). Thus, for a more comprehensive evaluation of QI methods that require a prior detection step, strategies to evaluate these methods on joint detection and quantification (JDQ) tasks are needed. Building on prior studies on evaluation techniques for such JDQ tasks[51,52], here we outline a framework to perform such evaluation.

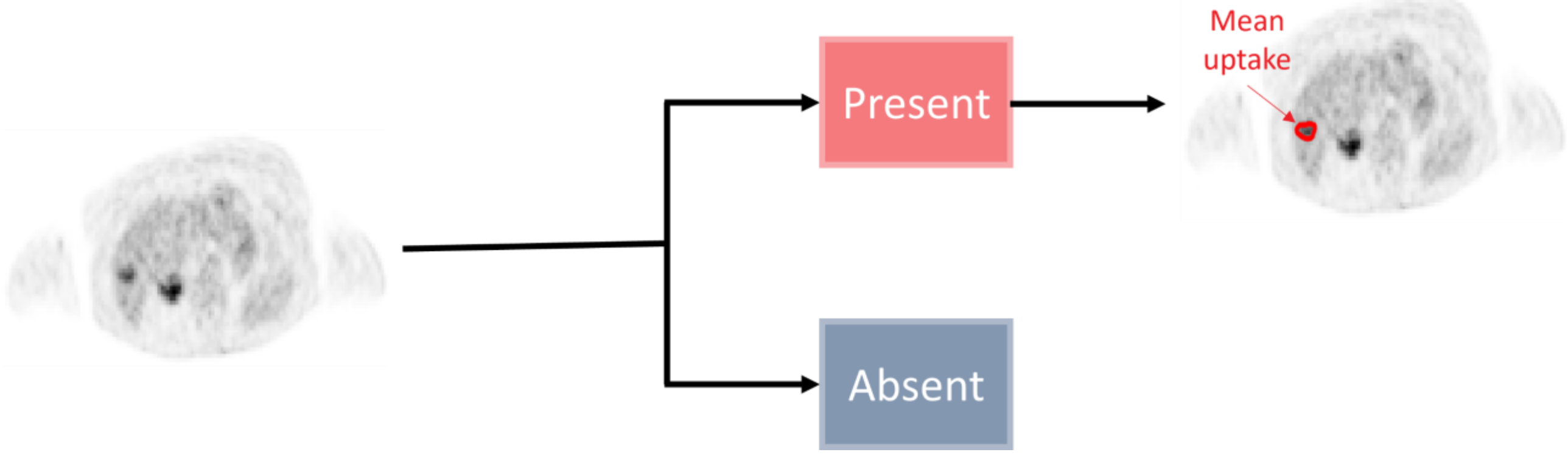

Fig. 5: A schematic demonstrating the process that an observer follows to decide on the tumor presence and then quantify the tumor when interpreting a PET image.

**Specification of the Clinical Task:** A JDQ task should involve performing both detection and quantification. Here, we refer to detection as a binary classification of signal presence that typically corresponds to abnormality, such as the tumor detection process demonstrated in Fig. 5. Quantification, as defined in the previous section, refers to the estimation of parameters such as MTV and TLG.

**Strategies to Perform Joint Detection and Quantification Task**: JDQ tasks can be performed sequentially or simultaneously. The sequential process typically involves first detection and then quantification, while the simultaneous process executes these tasks in one step.

In the sequential process, clinically, the detection step is typically performed by human observers. However, human-observer studies are time-consuming, tedious, and suffer from inter and intrareader variability. Given this challenge, anthropomorphic model observers (AMOs) can be used to identify promising methods for human-observer studies. These observers are designed to emulate human-observer performance in detection tasks[53,54]. One class of AMOs involves using anthropomorphic channels, which model the spatial-frequency selectivity of human visual system[55]. These are validated through psychophysical experiments by comparing their performance with human observers[55,56]. For QI methods developed for system and protocol optimization, model observers referred to as ideal observers have been developed, where an ideal observer is defined as one that yields the best possible performance on the detection task. For a thorough description of model observers, we refer the readers to He and colleagues[57].

Model observers have also been developed to perform the JDQ task simultaneously[58–60], including ideal observers for JDQ tasks[51]. Studies have applied anthropomorphic channels to model observers designed for JDQ tasks[58,60], which either have not been validated by human-observer studies or validated on other tasks, such as joint detection-localization task.58 Further validation of these observers on JDQ tasks is needed. Recently, an AI-based method has been used to approximate this observer[61]. Further, for JDQ tasks, Clarkson and colleagues[52] have shown that the maximum possible observer performance can also be expressed in terms of the Shannon information on individual task.

**FoMs for Evaluation**: Clarkson and colleagues[51] proposed an estimation receiver operating characteristic (EROC) curve that quantifies performance on the JDQ task. To plot the EROC curve, a utility score is defined to relatively quantify performance on the quantification task. Similar to a receiver operating characteristic (ROC) curve, the EROC curve is obtained by plotting the utility scores and false-positive fraction (FPF) at different thresholds. The area under the EROC curve (AEROC) provides a summary FoM for evaluations on the JDQ task. From a theoretical perspective, Wunderlich and colleagues[62] have provided a detailed description of various types of ROC curves using a utility-based formulation. In Fig. 6, we provide an example that uses this framework to estimate AEROC for the evaluation of two dose-reduction algorithms.

**Areas of Future Research**: Given the need for evaluation on JDQ tasks, continued research on developing and validating ideal and AMOs for JDQ task is needed. Here we note that in PET images reconstructed using MLEM and OSEM-based approaches, the noise in the reconstructed images have log-normal distributions with the noise covariance being dependent on the object[63]. Another important research area is the development and validation of AI-based model observers to perform JDQ tasks, expanding on recent advancements in this area[61]. Currently, NGSE methods are primarily developed for either detection or quantification tasks, and thus, new NGSE methods are needed for JDQ tasks.

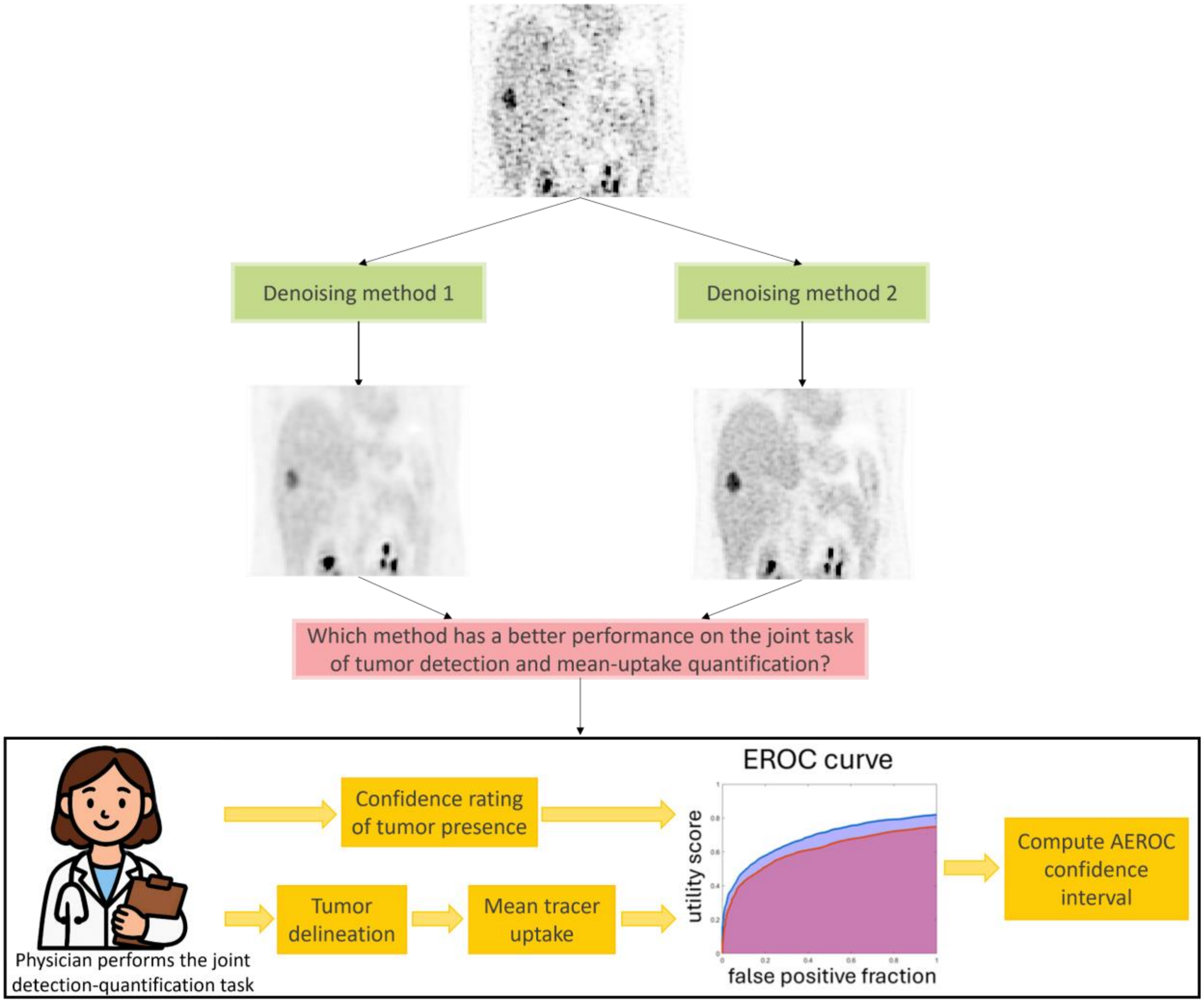

Fig. 6: A schematic of the JDQ task evaluation framework to evaluate two dose-reduction methods for PET on the task of tumor detection and estimation of mean uptake.

## V. Evaluation of QI Methods for Quantifying Multi-dimensional Parameters

Multidimensional parameters quantified from medical images, such as multiparametric quantitative imaging biomarkers and radiomic features (RFs), hold significant potential for disease staging and predicting treatment response[64,65]. Compared to single-parameter analyses, multi-dimensional parameters may offer a more comprehensive depiction of biological processes. However, the procedure to estimate these multidimensional parameters can impact their reliability. For example, the quantification of RFs from PET images involves several postprocessing procedures, including segmentation, filtering, and intensity binning, each of which introduces free parameters, the choice of which can substantially alter the value of the resulting RFs[66–68]. Thus, there is an important need for rigorous evaluation of QI methods that quantify multidimensional parameters.

Strategies to evaluate QI methods that quantify a single feature (e.g., MTV) may be insufficient when used to evaluate QI methods that yield multi-dimensional parameters. For example, consider the task of estimating a collection of RFs, where one QI method provides higher precision for certain RFs, whereas another QI method yields higher precision for the remaining RFs. This makes it challenging to compare the QI methods. To address this issue, we recognize that the eventual goal with using multi-dimensional parameters is clinical decision making, such as classifying tumors as malignant or benign and predicting therapy response. Thus, one possible evaluation approach is assessing the accuracy of these methods on making this clinical decision.

Typical research studies using muti-dimensional parameters for clinical-decision making involve first developing and training a model based on machine learning algorithms, such as logistic regression, support vector machine, random forest and neural networks[69]. For developing prognostic models, statistical methods such as the Cox proportional hazard ratio and Kaplan-Meier estimator are used. We assume that such a model has been already developed for each QI method being evaluated and, in this article, our focus is on evaluation of the QI methods. For this purpose, we outline a framework below, as presented in Fig.7. The structure of the framework is guided by the methodology for clinical evaluation proposed in RELAINCE guidelines[70].

**Specification of the Clinical Task:** Multidimensional parameters are typically used as inputs to diagnostic models, such as classifying lesions as malignant or benign and cancer staging; prognostic models, such as predicting cancer recurrence risks and overall survival; and predictive models, such as predicting treatment response.

**Study-Type Selection:** These studies can be retrospective, prospective, or real-world evaluation studies[70]. Depending on factors such as risk, clinical-data availability, and level of evidence required, the study-type determination can be made.

**Data Collection:** QI methods should be evaluated using test data that was not used during model training or validation[71]. For comparing performance over a certain target population, the collected test data should be representative of this population[70]. In retrospective studies where the database has been defined, patients should be randomly selected to avoid selection bias. The data size depends on the specific performance claim being tested. Power analysis, guided by pilot or earlier studies, can help determine the required sample size.

**Defining a Reference Standard:** For diagnostic tasks, external standards, for example, derived from invasive means such as biopsy pathology, can be used if available[72]. Additionally, expert panels can provide reference standards for disease diagnosis[73]. For prognostic and predictive tasks, outcomes such as overall survival and progression-free survival can serve as reference standards[70].

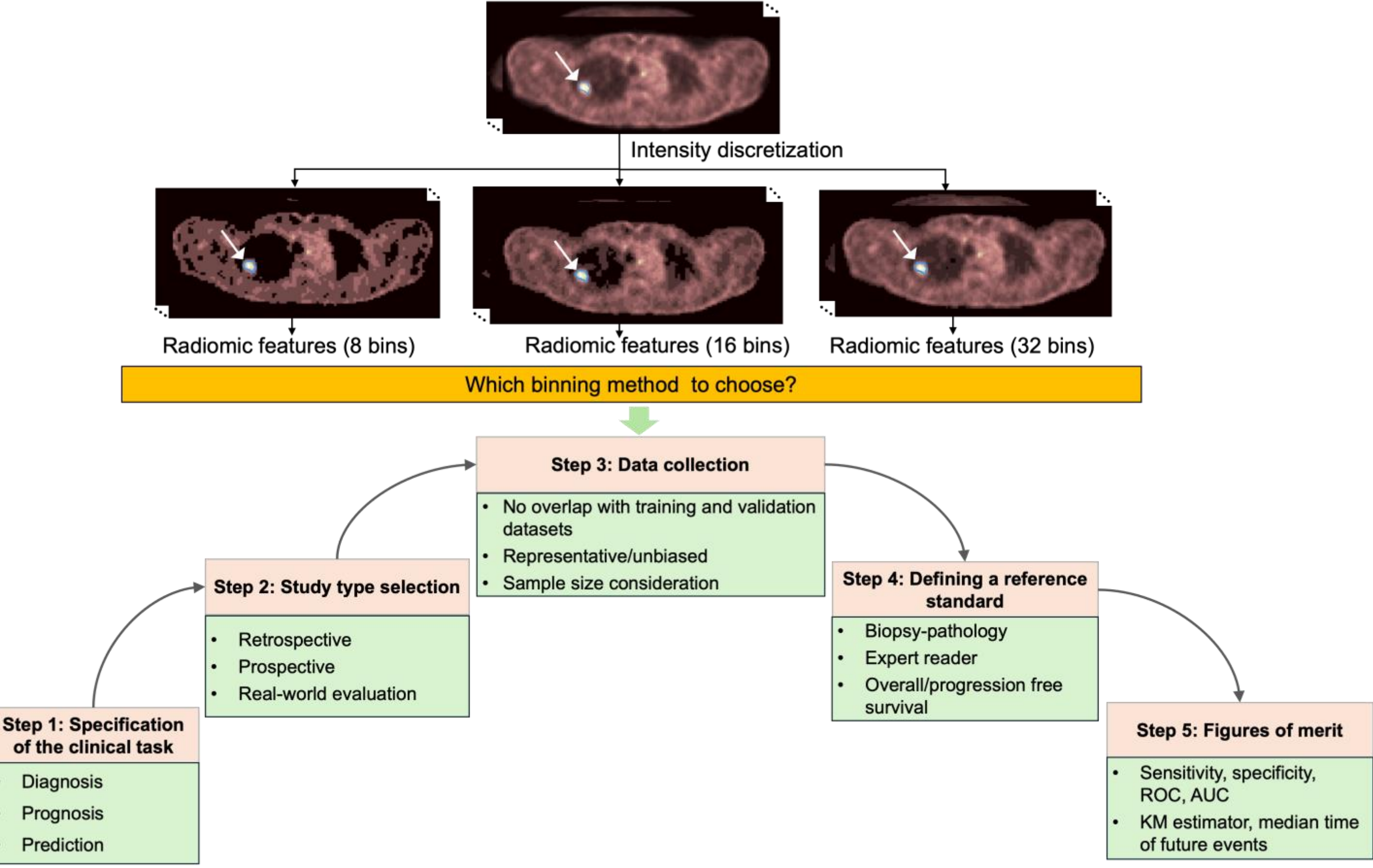

Fig. 7: The framework for evaluating QI methods in multi-dimensional feature quantification, illustrated by evaluating different numbers of bins used for intensity discretization to compute radiomic features. The white arrow indicates the lesion.

**FoMs for Evaluation:** FoM should be selected based on the clinical task. A summary of the FoMs is provided previously[70]. Here, we present the key evaluation criteria and corresponding FoMs relevant to different clinical applications.

- *Diagnostic task:* Diagnostic tasks are typically framed as classification problems, the FoMs of sensitivity, specificity, the ROC curve, and area under the ROC curve (AUC) can be used. Confidence intervals can be used to quantify uncertainty of the FoM.
- *Prognostic and predictive task:* Prognostic and predictive tasks can also be framed as classification problems, for example, classifying patients into responders and non-responders. In such cases, similar FoMs for diagnostic tasks can be used. For time-to-event predictions, appropriate FoMs include Kaplan–Meier estimators and median time of future events.

**Areas of Future Research:** The outlined framework, while providing a mechanism to assess the QI method in the context of clinical decision making, does not assess performance on the quantification task. The use of VITs may provide a mechanism to measure performance on quantitative tasks, and this is an important research frontier. Relatedly, a future research direction is developing FoMs that quantify performance of QI methods in estimating multidimensional parameters.

A major area in PET where multidimensional parameters are estimated is dynamic PET. Clinical expansion of dynamic imaging as a fully quantitative approach has gained significant attention[74]. In dynamic imaging, macro-kinetic parameters, using graphical analysis methods such as Patlak and Logan, and micro-kinetic parameters, such as those from multi-tissue compartment kinetic models, are multidimensional parameters. However, the development of evaluation methods for evaluating QI methods that estimate these parameters is still evolving[74]. VITs may provide a mechanism to evaluate QI methods for estimating kinetic parameters with a known ground truth. The framework outlined here could be used provided a clear clinical decision-making task can be identified. Nevertheless, research is needed for objective evaluation of QI methods for dynamic PET, ensuring accurate estimation of kinetic parameters.

## VI. Discussion

The article outlines four emerging frameworks for evaluating QI methods, covering the scope of conducting evaluation in both virtual and clinical settings for quantifying single-dimensional and multidimensional parameters, including scenarios without gold standards. These frameworks can assist with assessing performance and supporting technological advancements in quantitative imaging, including designing and optimizing new systems and protocols, and methods for image reconstruction, enhancement, and analysis.

The article focuses on digital simulation and clinical data-based evaluation strategies. Another important mechanism to perform evaluations is with physical phantoms, which offer the advantage of evaluating new QI methods with known ground truth and on physical scanners. An important area of active research is designing physical phantoms to model the human physiology and anatomy more realistically and simulate population variability. With recent advancement in 3D-printing technology, phantoms that model human anatomy, referred to as anthropomorphic phantoms, are being developed[75,76]. Further, phantoms that model respiratory motion[77,78]. heterogeneous soft-tissue texture[79,80], and tumor heterogeneity[79,81] are also being developed.

Another future research direction in the era of AI is the development of frameworks to evaluate human-in-the-loop AI approaches that yield quantitative values. Such efforts refer to collaboration between human (e.g., physician) and AI algorithms during the learning process[82]. An important question is the objective evaluation of such algorithms as they are being continuously updated. Strategies to evaluate continuously learning AI-based algorithms[70] can be adapted for conducting these evaluations. Such a question can also be contextualized broadly in the field of implementation science and knowledge translation, where an algorithm should be evaluated not merely based on FoMs perceived to be important by algorithm creators, but by algorithm users themselves (e.g., physicians) to overcome barriers related to the use of the developed solutions and to enable their routine adoption; especially algorithms should be assessed according to four-core requirement to provide: reason, means, method, and desire to use by physicians[83].

The frameworks outlined in this article are designed to evaluate QI methods at a population level. However, there may also be a need to evaluate QI methods at a per-patient level, such as to assess interchangeability of a QI method to standard of care for individual patients or to optimize a method for a specific patient[28,84]. One statistical measure used for assessing interchangeability is the individual equivalence index[85,86], which compares the average squared difference between measurements obtained from the QI method and the standard of care to the average squared difference observed when the standard of care is used at two separate occasions. This method has been used to assess the interchangeability of CT and MR imaging in patients with femoroacetabular impingement for estimating acetabular version[86]. To enable personalized evaluation in virtual settings, the concept of digital twinning can be considered. Digital twins are virtual avatars that are created and personalized for individual patient[87,88]. These twins, as a reference standard for the patient, can enable the evaluation of QI methods at a per-patient level. In this context, when findings in virtual settings are directly applied to real patients, VVUQ can be considered for quality assurance.

In summary, based on studies in the literature, we have presented four emerging evaluation frameworks for objective evaluation of QI methods, describing their applications, strengths, and areas for advancement. We recommend that when these frameworks are used for evaluation, and particularly when they are used to evaluate AI algorithms, an appropriate claim be generated to report the performance of the QI method as per the RELAINCE guidelines. Our vision is that these evaluation frameworks can facilitate translation of QI methods ultimately contributing to improvements in quality health care and better treatment outcomes.

## Clinical Care Points

• In quantitative medical imaging, numerical or statistical features are estimated from medical images for clinical decision making. The field holds significant potential across diverse clinical applications.

• For clinical translation and to gain wider clinical usage, validation of quantitative imaging

methods on clinical tasks is crucial.

• We outline four emerging frameworks to perform such validations.

• Our vision is that these evaluation frame- works will contribute to improved clinical decision making with quantitative imaging.

# Declaration of AI and AI-assisted technologies in the writing process

During the preparation of this work the author(s) used ChatGPT to generate the cartoon for PET scan in Fig. 2 and the physician in Fig. 6. After using this tool/service, the author(s) reviewed and edited the content as needed and take(s) full responsibility for the content of the publication.

# Acknowledgement

The authors would like to thank Kyle Myers, PhD, William Paul Segars, PhD, Ehsan Abadi, PhD, and Guobao Wang, PhD for helpful discussions on this manuscript.